\documentclass[preprint2]{aastex}

\shorttitle{Intermittency out of the Ecliptic Plane}
\shortauthors{Wawrzaszek et al.}

\usepackage{graphicx}
\usepackage{natbib}
\usepackage{amssymb}
\usepackage{amsmath} 
\usepackage{amsfonts}

\begin{document}


\title{Evolution of Intermittency in the Slow and Fast Solar Wind Beyond the Ecliptic Plane}

\author{A. Wawrzaszek\altaffilmark{1}, M. Echim\altaffilmark{2,3}, W.~M. Macek\altaffilmark{1,4},   \and R. Bruno\altaffilmark{5}}
 
\altaffiltext{1}{Space Research Centre, Polish Academy of Sciences, Warsaw, Poland \email{anna.wawrzaszek@cbk.waw.pl}}

\altaffiltext{2}{The Belgian Institute for Space Aeronomy, Brussels, Belgium \email{marius.echim@oma.be}}

\altaffiltext{3}{Institute of Space Science, M\u{a}gurele, Romania}

\altaffiltext{4}{Faculty of Mathematics and Natural Sciences, 
Cardinal Stefan Wyszy\'{n}ski University, Warsaw, Poland \email{macek@cbk.waw.pl}}

\altaffiltext{5}{Institute for Space Astrophysics and Planetology, Roma, Italy \email{roberto.bruno@iaps.inaf.it}}

\begin{abstract}

We study intermittency as a departure from self-similarity of the solar wind magnetic turbulence and investigate
the evolution with the heliocentric distance and latitude. We use data from the \textit{Ulysses} spacecraft measured during
two solar minima (1997-1998 and 2007-2008) and one solar maximum (1999-2001). In particular, by modeling a
multifractal spectrum, we revealed the intermittent character of turbulence in the small-scale fluctuations of the
magnetic field embedded in the slow and fast solar wind. Generally, at small distances from the Sun, in both the
slow and fast solar wind, we observe the high degree of multifractality (intermittency) that decreases somewhat
slowly with distance and slowly with latitude. The obtained results seem to suggest that generally intermittency in
the solar wind has a solar origin. However, the fast and slow streams, shocks, and other nonlinear interactions can
only be considered as the drivers of the intermittent turbulence. It seems that analysis shows that turbulence beyond
the ecliptic plane evolves too slowly to maintain the intermittency with the distance and latitude. Moreover, we
confirm that the multifractality and intermittency are at a lower level than in the ecliptic, as well as the existence of
symmetry with respect to the ecliptic plane, suggesting that there are similar turbulent properties observed in the
two hemispheres.

\end{abstract}


\section{Introduction}
\label{sec:w:int}
 
Starting with the pioneering work of Burlaga (\citeyear{Bur91a}) a considerable number of studies have been dedicated to 
the analysis of the intermittent nature of the solar wind turbulence (see \cite{BruCar13} for a review). 
In particular, the phenomenon of intermittency, as the inhomogeneity of the energy transfer rate in the turbulent nonlinear
cascade, has been identified in fluctuations of plasma variables and of the magnetic field in the entire heliosphere \citep{MarLiu93,Bur95,PagBal02,Bruet03}, in the distant heliosheath \citep{BurNes10,Macet11,Macet12}, and near the heliopause \citep{Macet14}. 
The level of intermittency in the solar wind turbulence has been mainly determined by the analysis of the anomalous scaling of structure functions \citep{MarLiu93}, by considering probability distribution functions (PDFs) of fluctuations and departure from Gaussianity for smaller scales \citep{Bruet03,Yoret09}, by using the scale behavior of the flatness (or fourth-order moment) and fitting of the PDFs \citep{YangTam2010}, or by fitting models of turbulence \citep{PagBal02}.\\
Multifractal analysis is a higher-order method used to reveal the intermittent nature of various characteristics of turbulence (see,
for instance, \citep{Bur91b,Buret03,WawMac10}). In particular, the degree of multifractality is
a measure of the level of heterogeneity in the local scaling indices; in other words, it informs about the departure from a strict self-similarity \citep[e.g.,][]{Fri95}. 
Thus, it is also a quantitative measure of the intermittency successfully used in many studies of turbulence \citep{MacWaw09,BurNes10,WawMac10,Macet12,Macet14}. \\
The analysis of observations at the ecliptic showed that fast solar wind is generally less intermittent than slow wind for both
wind speed and magnetic field components \citep{MarLiu93,Bruet03}. Moreover, \cite{Bruet03} and \cite{YangTam2010} studied the
intermittent properties of the fast wind in the inner heliosphere from Helios data (between $0.3$ and $0.9$ AU) 
and concluded that fast wind intermittency increases with the distance from the Sun.\\
Considerable  effort  has  been  devoted  to  investigating  the evolution of intermittency in interplanetary magnetic fields 
with distance and latitude in the solar wind \citep[e.g.,][]{PagBal02,PagBal03,Yoret09}.
In particular, \cite{PagBal02} focused on the different intermittent levels of magnetic field fluctuations 
measured by \textit{Ulysses} during solar minimum (1994-1996) and solar maximum (2000-2001) and found a high level 
of intermittency throughout the both phases of the solar cycle. Moreover, they showed that the slow wind has a lower level 
of intermittency as compared with the fast flow. 
However, these authors used only the wind speed as a criterion to discriminate between pure coronal fast and typical equatorial slow wind. 
Further analysis of pure polar coronal fast wind at solar minimum between 1994 and 1996 \citep{PagBal03} shows that inertial range intermittency at time scales between 40 and 200 seconds, described as departure of PDFs from Gaussianity, increases with increasing radial distance from the Sun. \\
\cite{Yoret09} performed a detailed selection of the so-called ``pure" states of 
the solar wind encountered by \textit{Ulysses} between 1992 and 1997. It was shown that slow wind is more intermittent than fast wind and no radial evolution was found for the slow wind. However, the heliocentric radial variations during the time intervals studied by \cite{Yoret09} were rather limited ($5.1-5.4$ AU), as was the heliolatitude range ($L<20^{\circ} $).
There are very few comparative studies of solar wind  intermittent turbulence at solar minimum and solar maximum out of the
ecliptic plane. Therefore, a systematic analysis of data for a larger range of radial distances and heliolatitudes may allow 
a better understanding of the nature of intermittency out of the ecliptic plane.\\
In this study, we use \textit{Ulysses} data recorded during the last two minima of the solar cycle (1997-1998 and 2007-2008), as well as
during the previous maximum (1999-2001); the database includes 130 time intervals of ``pure" slow and fast solar wind,
whose selection procedure is described in the next section. Based on these data, we construct a more complete picture of the multifractal scaling and intermittent nature of the turbulence beyond the ecliptic plane. 
This is the comprehensive study of such a magnitude devoted to the multifractal spectrum of fluctuations of 
the interplanetary magnetic field strength measured at different heliocentric distances and heliographic latitudes 
during two solar minima (1997-1998, 2007-2008) and a solar maximum (1999-2001). 
The selected data cover a radial range between 1.5 and 5.4 AU and
a heliolatitude range between $-80^\circ$ and  $70^\circ $.

\section{\textit{Ulysses} Data}
\label{sec:w:ud}

\textit{Ulysses} is a joint ESA-NASA mission launched in 1990 October. The probe completed roughly three solar orbits before the mission ended in 2009 June.  
The orbit of \textit{Ulysses} was periodic (6.2 years), with perihelion at 1.3 AU and aphelion at 5.4 AU and a latitudinal excursion of $\pm 82^{\circ}$ allowing the study of both latitudinal and radial dependence of the solar wind over a wide range of solar activity conditions \citep{Smiet95}. In this study, we focused on the slow and fast solar wind data measured in situ by the \textit{Ulysses} during two solar minima (1997-1998, 2007-2008) and a solar maximum (1999-2001). To identify the solar wind type and origin we define a set of criteria and thresholds applied on plasma and magnetic field data. Thus, we use five parameters to discriminate between the fast and slow solar wind state: (1) the radial velocity, (2) the proton density, (3) proton temperature parameters measured by the SWOOPS instrument \citep{Bamet92}, (4) the oxygen ion ratio ($\mathrm{O}^{+7}/\mathrm{O}^{+6}$) from SWICS \citep{Gloet92}, and (5) the magnetic compressibility factor calculated from magnetic field measurements by the VHM-FGM magnetometer \citep{Balet92}. For each of the five parameters, we defined a threshold value as a transition from slow to fast wind:(1) $V_{\mathrm{thr}}=450$ km s$^{-1}$, (2) $n_{\mathrm{thr}}=0.2$ cm$^{-3}$, (3)$T_{\mathrm{thr}}=4*10^{4}$ K, (4) $(\mathrm{O}^{+7}/\mathrm{O}^{+6})_{\mathrm{thr}}= 0.1$, and (5) ${C_B}_{\mathrm{thr}}=0.1$  \citep{vonet01,Yoret09,BruCar13,Lepet13}.
We also take into account the solar cycle variation of these thresholds \citep[e.g.,][]{Kaset12} and exclude time intervals corresponding to interplanetary transients like shocks and CMEs, identified by Gosling and Forysth (see http://www.sp.ph.ic.ac.uk/Ulysses/shocklist.txt) and Gosling and Ebert (see http://swoops.lanl.gov\\/cme\_list.html). 
We identified $93$ time intervals of fast solar wind ($38$ during solar maximum and $55$ during solar minimum) and $37$ time intervals of slow solar wind ($28$ during solar maximum and $9$ during solar minimum). The multifractality is checked for each data set that contains a few days long time series of the magnetic field strength measured with a $0.5$ Hz sampling rate. 

\section{Multifractal Analysis}
\label{sec:w:ma}

The multifractal approach is a generalization of fractal analysis used to characterize objects that demonstrate a variety of self-similarities at various scales.  A detailed description of the multifractal theory can be found in , e.g., the works of \citet{Halet86} or \citet{Fal90}. One of the basic characteristics of multifractal scaling is the multifractal spectrum $f(\alpha)$, sketched in Figure~1 of the paper by \cite{WawMac10}. As a function of a singularity strength $\alpha$, $f(\alpha)$ portrays the variability in the scaling properties of the measures. In particular, the width of the spectrum quantifies the degree of multifractality of a given system \citep[e.g.,][]{Ott93, MacWaw09}.\\
 There are several techniques to obtain  multifractal spectra from the experimental data. In this study, we use two methods: 
the Partition Function technique \citep{Halet86} and the direct method of determination of a multifractal spectrum proposed by \citet{ChhJen89},
 both successfully tested in many situations for space plasma turbulence \citep{MacSzc08,BurNes10,Lamet10} and briefly described below. \\
In the first step of the analysis, we construct a multifractal measure \citep{Man89} from the first moment of increments of magnetic fluctuations at a scale $l$,
\begin{equation}
\epsilon(x_i,l)=\mid B(x_i+l)-B(x_i )\mid  
\end{equation}
where $i = 1, \ldots, N = 2^{n}$, with $n$ from $17$ to $19$ denoting the data set length as a power of 2, while $B$ denotes the magnetic field strength. Next, we decompose the signal in segments of size $l$ and associate with each $i$th segment 
a probability measure defined by
\begin{equation}
p(x_i,l)\equiv \frac{\epsilon(x_i,l)}{\sum_{i=1}^{N}\epsilon(x_i,l)}=p_i(l).
\end{equation}
This quantity can be interpreted as a probability that the portion of fluctuation is transferred to a segment of size $l$. As is typical, at a given position $x=v_{\mathrm{sw}} t$, 
where $v_{\mathrm{sw}}$ is the average solar wind speed, the temporal scales measured in units of sampling time $\Delta t$ can be interpreted as 
the spatial scales  $l=v_{\mathrm{sw}} \Delta t$ (the Taylor hypothesis).
Next, the scaling of the Partition function, $\chi (q,l)=\sum_{i=1}^{N(l)}(p_{i} (l))^q $,
 for  various box size $l$ and moments of order $q$ is considered according to
\begin{equation}
\chi (q,l)\propto l^{\tau(q)},
\label{e:w:spf}
\end{equation}
where $\tau(q)$ is the scaling exponent.  The scaling exponents $\tau(q)$ are extracted from the slopes of the logarithm of the partition 
function $\chi (q,l)$ versus $\log l$ (Eq. (\ref{e:w:spf})). 
In the next step, $f(\alpha)$ is determined by a Legendre transformation of the $\tau(q)$ curve 
with $\alpha(q)=  \frac{d}{dq}[\tau(q)]= \tau'(q)$ and $f(\alpha(q))= q\alpha(q)-\tau(q)$ \citep{Halet86,ChhJen89}.
In the case of the direct method of determination of a multifractal spectrum, the so-called pseudoprobability measure, $\mu_{i}(q, l)$,  
is determined from
\begin{equation}
\mu_{i}(q, l) \equiv \frac{p^{q}_{i}(l)}{\sum_{i=1}^{N} p^{q}_{i}(l)}.
\label{e:ms:mui}
\end{equation}
Next, the multifractal spectrum  
can be calculated directly from the average of 
the pseudoprobability measure $\mu_i(q, l)$ in Eq.~(7) \citep{ChhJen89,Chhet89}: 
\begin{equation}  
f(q)= \lim_{l\to{0}}~\frac{\langle \log \mu_i(q, l) \rangle}{\log (l)}.
\label{e:ms:fqK}
\end{equation}
where $\langle \ldots \rangle$ indicates ensemble averaging.
The corresponding average value of the singularity strength is given by\citep{Chhet89}  
\begin{equation}  
\alpha(q)= \lim_{l\to{0}}~\frac{\langle \log p_i(l) \rangle}{\log (l)}. 
\label{e:ms:aqS}
\end{equation}

The scaling range where a linear regression is applied to determine $\tau(q)$, $f(q)$, and $\alpha(q)$ is not known a priori, and it must be chosen before the analysis. 
In this study, we apply an automatic selection procedure proposed by \cite{SauMul99}. 
In most cases, we identified a scaling range from 4 s to a few days that can be somewhat related to the existence of the inertial range and with the dissipation effects observed on smaller scales.
Cases, for which the optimal scaling range could not be determined, have been rejected from a further analysis.
\\
The two points $\alpha_{\mathrm{min}}$ and $\alpha_{\mathrm{max}}$, at which $f (\alpha) = 0$,
quantify  the degree of multifractality \citep{MacWaw09,WawMac10,Macet14}:
\begin {equation}
\Delta\equiv \alpha_{\mathrm{max}} -\alpha_{\mathrm{min}}.
\label{e:w:dm}
\end{equation}
Owing to the limited amount of data available from \textit{Ulysses} in the solar wind, we can only determine the points near the maximum of $f(\alpha)$. Therefore, in the last step of the analysis we fit the observations with a theoretical two-scale binomial multiplicative model \citep{Buret93,MacSzc08} to extrapolate the values of $\alpha_{\mathrm{max}}$  and $\alpha_{\mathrm{min}}$ and to obtain the degree of multifractality.\\
 It is worth to noting that the agreement of the multifractal spectrum with the phenomenological intermittent two-scale model confirms that multifractal structures detected at small scales are generated by an intermittent cascade. Moreover, the degree of multifractality calculated by using fitted two-scale model parameters ($\Delta=\bigg{|} \frac{\log (1 - p)}{\log l_2} - \frac{\log (p)}{\log l_1} \bigg{|}$; for more details, see, e.g. \cite{MacWaw09}) measures inhomogeneity of the considered fluctuations and their deviation from a strict self-similarity. That is why $\Delta$ is also a measure of intermittency, which is in contrast to self-similarity \citep[ch. 8]{Fri95}.

\section{Results}
\label{sec:wa:res}
\begin{figure*}[!h]
\centering
\includegraphics[scale=0.71]{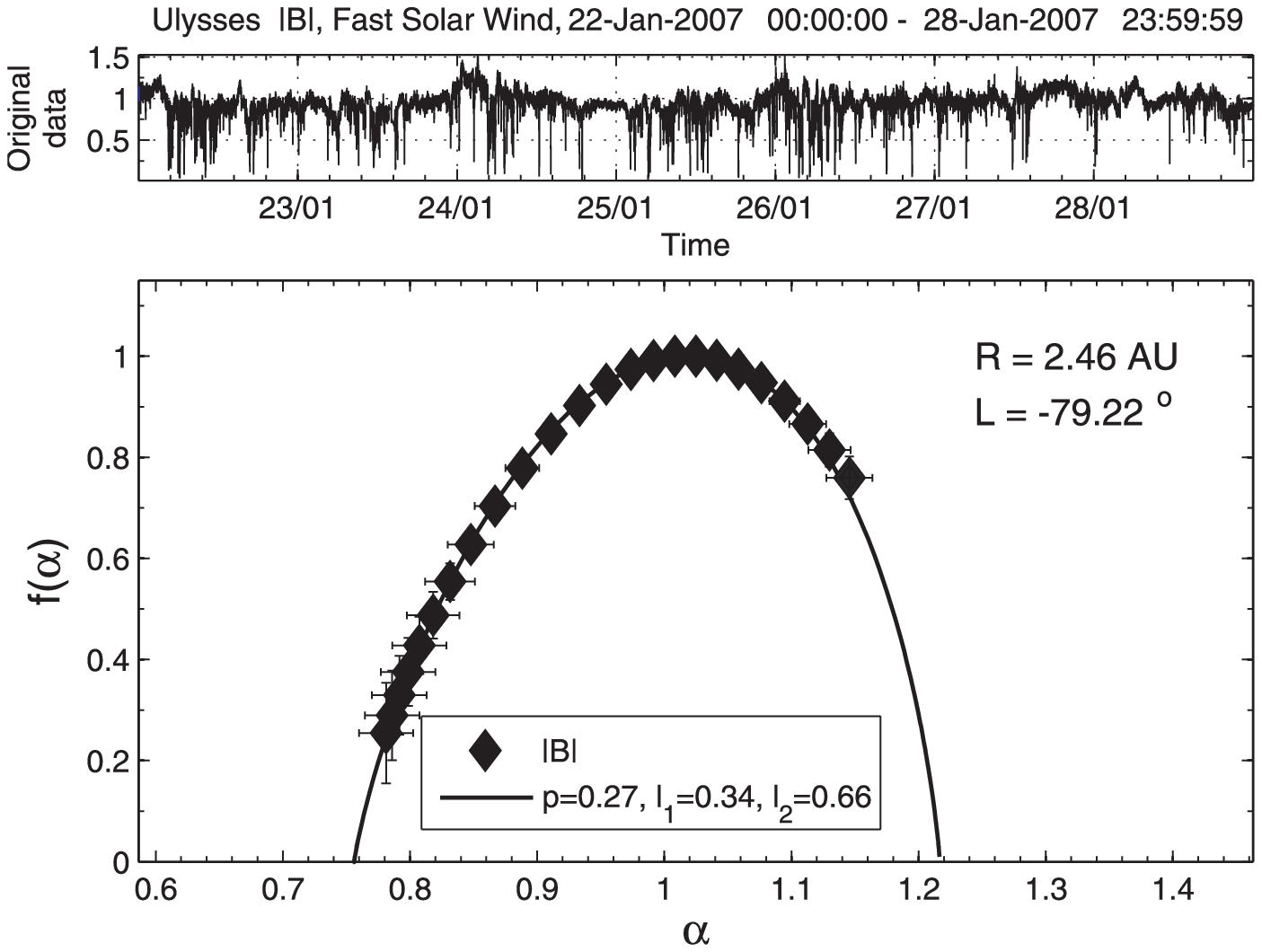}
 \caption{Multifractal spectrum $f(\alpha)$ as a function of singularity strength $\alpha$ (diamonds) determined for the magnetic field strength of the fast solar wind measured by \textit{Ulysses} in 2007 at $2.46$ AU, $-79.22^{\circ}$. Continuous line shows a theoretical two-scale model fitted with observations.}
 \label{f:w:sp}
\end{figure*}

Figure ~\ref{f:w:sp} presents an example of an $f(\alpha)$ multifractal spectrum obtained using the experimental values of the magnetic field  strength for the fast solar wind measured by \textit{Ulysses} in 2007 (from January 22 to 28) at $2.46$ AU and $-79.22^\circ$. 
The continuous line shows the $f(\alpha)$ spectrum derived from the theoretical two-scale model \citep{MacSzc08}
fitted with the observations; the theoretical fit allows us to determine the width of the multifractal spectrum as the measure of multifractality (intermittency), $\Delta=0.46\pm 0.04$, as defined by Eq. \ref{e:w:dm} and discussed in the previous paragraph.

 \begin{figure*}[!h]
\centering
\includegraphics[scale=0.725]{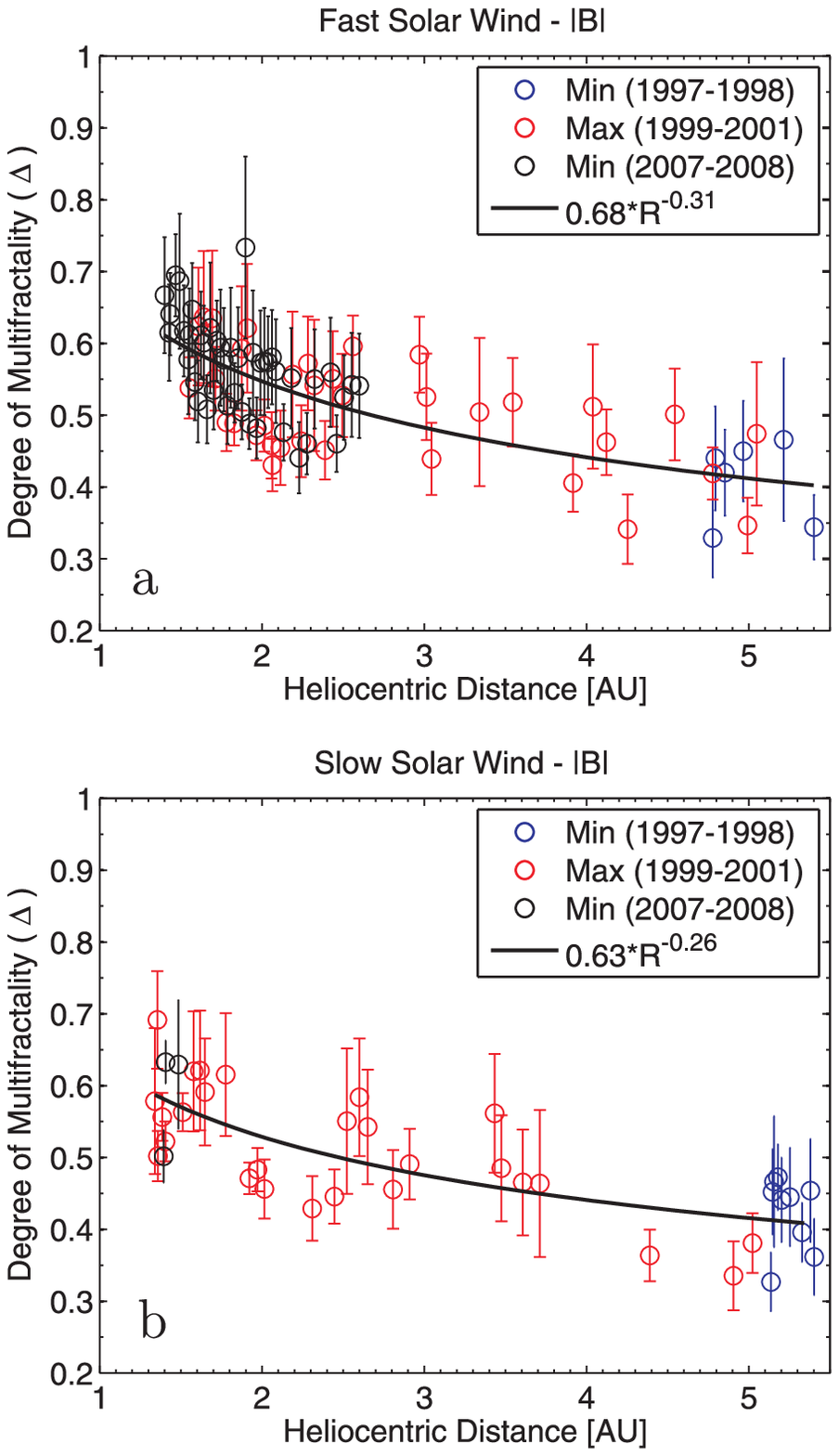}
 \caption{Degree of multifractality as a measure of intermittency
in the magnetic field strength in dependence on distances from the Sun
during two solar minima (1997-1998 and 2007-2008) and a solar maximum (1999-2001), respectively. Power law fit $\Delta=aR^{b}$ to data shown by a continuous line is also indicated.}
 \label{f:w:rad}
 \end{figure*}

\begin{figure*}[!h]
 \centering
\includegraphics[scale=0.725]{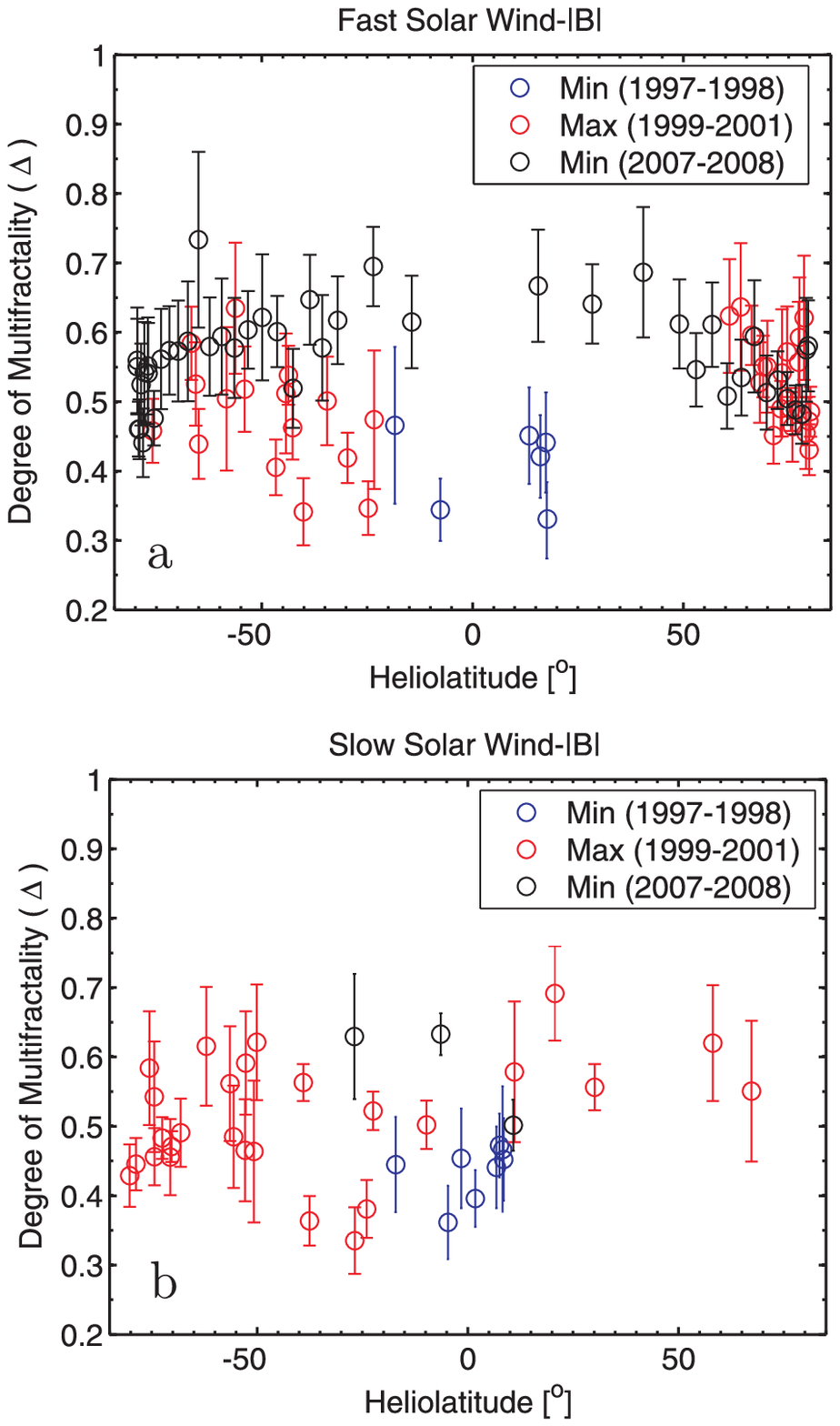}
 \caption{Degree of multifractality as a measure of intermittency
in the magnetic field strength field in dependence on the heliographic latitude during two solar minima (1997-1998 and 2007-2008) and a solar maximum (1999-2001), respectively.}
 \label{f:w:lat}
 \end{figure*}

\begin{figure*}[!h]
 \centering
\includegraphics[scale=0.65]{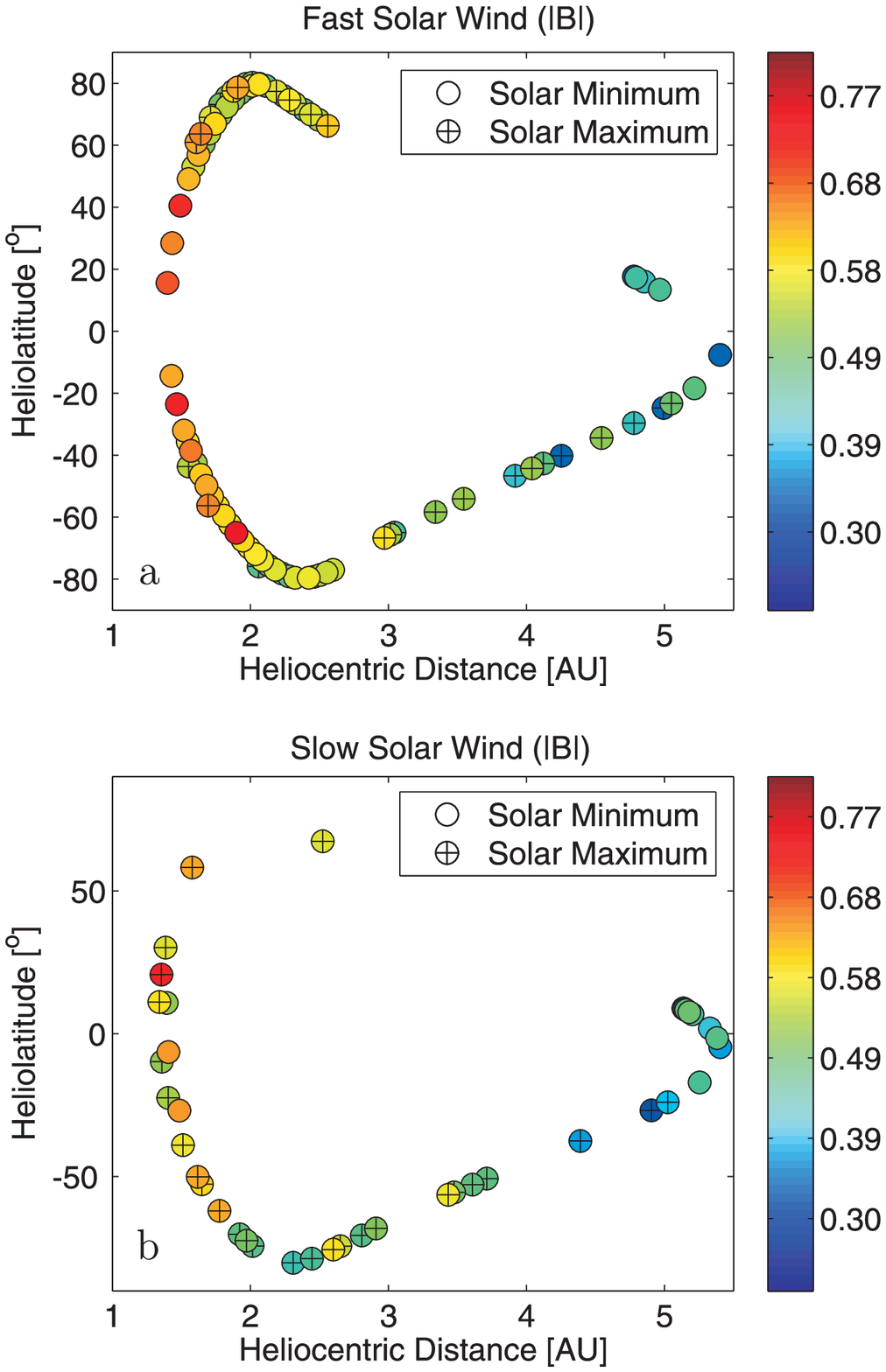}
 \caption{Map of the degree of multifractality as a measure of intermittency determined for fast (a) and slow (b) solar wind during two solar minima (1997-1998 and 2007-2008) and a solar maximum (1999-2001), respectively. Color denotes the values of the parameter $\Delta$ determined for data at different heliocentric distances and heliographic latitudes.}
 \label{f:w:fast_slow}
 \end{figure*}

The parameter $\Delta$ was computed for all selected time intervals during two solar minima (1997-1998, 2007-2008) and  maximum (1999-2001).
The results are shown in Figures~\ref{f:w:rad} and ~\ref{f:w:lat}. \\
Figure~\ref{f:w:rad} suggests a slow evolution of intermittency in the fast and slow solar wind with the distance $R$ from the Sun. 
The fast wind multifractal spectra seem to exhibit a small decrease of intermittency with the distance as $\Delta \propto  R^{-0.31\pm 0.03}$. The rate of this decrease has been determined by the least squares linear fit to the log-log values of $\Delta$ against $R$ to all points both from solar minimum and maximum (Figure \ref{f:w:rad}a). 
For the slow solar wind (Figure \ref{f:w:rad}b), the evolution of intermittency with distance $R$ is more variable but anyway the decrease with distance with rate $-0.26\pm 0.04$ is observed anyway. \\
A latitudinal dependence of the multifractality of turbulence is also found, as illustrated in Figure~\ref{f:w:lat}. 
 The fast solar wind at solar minimum exhibits a decrease in intermittency as the latitude increases; the smallest values of multifractality are found near solar pole. 
Another significant effect is the existence of a symmetry with respect to the ecliptic plane, observed mainly during solar minima (1997-1998 and 2007-2008), which confirms 
previous observations \citep{Bavet00,WawMac10}. This is an indication that the fast polar solar wind has similar turbulent properties 
in the two hemispheres. We cannot confirm this trend at solar maximum since the statistics of slow wind intervals in our database is rather poor and the solar wind is much more
variable.\\
The results denoted in Figures~\ref{f:w:rad} and~\ref{f:w:lat} by blue circles correspond to 
solar minimum conditions (1997-1998), heliocentric distances of the order of $\sim 5$ AU, and heliographic latitudes smaller than $20^{\circ }$. 
For this data subset, we observe that the level of multifractality is slightly lower for fast 
than for slow solar wind, with mean  $\Delta$ about $0.40$ (compared to $0.44$ for the slow wind). 
These results are consistent with previous studies performed in the ecliptic plane \citep{MarLiu93,Bruet03} and 
with previous analyses of \textit{Ulysses} data for the years 1992-1997 \citep{Yoret09}, which postulate that intermittency in the slow solar wind is higher than for the fast wind. \\
Results for the second minimum (2007-2008), marked in Figures~\ref{f:w:rad} and ~\ref{f:w:lat} by black circles, seem to suggest somewhat different behavior. 
However, the data from this time interval have been collected for a wider range of 
latitudes, $-80^{\circ} \le L \le 80^{\circ}$, and for a radial distance from the Sun between $1.4$ and $2.6$ AU.
Our hypothesis here is that the effect of change on solar wind multifractality with heliographic latitude
may interfere with variations of this parameter observed with
respect to the distance from the Sun.
The statistics of ``pure'' slow wind data is very poor (only three intervals for 2007-2008), correlated with the
end of Ulysses mission; therefore, a comparison of the variation of multifractality between fast/slow wind and the solar cycle is not possible.\\
In general, in the case of \textit{Ulysses} data, it is difficult to separate latitudinal and radial evolution of intermittency. The calculated multifractality as function of both heliocentric distance and heliographic latitude for the fast and slow solar wind are summarized in Figures \ref{f:w:fast_slow}a and \ref{f:w:fast_slow}b, respectively. 
One may see that fluctuations of the magnetic field strength both in the fast and slow solar wind exhibit increased
intermittent features at small distances from the Sun. 
Moreover, once again we see that at small heliolatitudes and  at distance of about $5$ AU, the slow wind is more intermittent than the fast. Further, analysis of results presented in Figures~\ref{f:w:rad} and~\ref{f:w:lat} shows a rather similar range of the degree of intermittency determined for the solar maximum and minimum. In other words, we have less dependence on the phases of solar activity than on the type of the solar wind \citep{PagBal02}.\\
Additionally, the multifractal analysis of \textit{Ulysses} magnetic field data presents 
a somewhat smaller degree of multifractality, $\Delta=0.35-0.73$, than  those obtained previously for velocity fluctuations, $\Delta=1.34-1.52$ \citep{WawMac10}.\\

\section{Conclusions}
\label{sec:wa:con}

In this Letter, we have performed systematic studies of the multifractal and intermittent behavior 
of fluctuations of the interplanetary magnetic field strength for a broad range of heliographic latitudes, 
heliocentric distances, and for the maximum and minimum phase of the solar cycle. 
The obtained results can be summarized as follows: 
(1) the slow solar wind  during solar minimum (1997-1998), at distances of $\sim 5$ AU and close 
to the equatorial plane, exhibits on average a higher level of intermittency than fast solar wind;
(2) the fast solar wind measured during the second minimum (2007-2008), at distances of $1.4-2.6$ AU and latitudes of $-80^{\circ}\div 80^{\circ}$ reveals a decrease of intermittency;
(3) the level of intermittency during solar maximum present non-orderly behavior; however, in many cases, we observe the same degree as those determined for data from solar minimum; and (4) during solar minima (1997-1998 and 2007-2008), similar intermittent properties of the fast polar solar wind in the two hemispheres are observed.\\
Generally, at small distances from the Sun, both in the slow and fast solar wind, we observe the highest degree of multifractality, 
suggesting that most of the intermittent events have a solar origin. Moreover, the level of intermittency seems to decrease also with latitude.
This effect can be related to the fact that at radial distances of the order of 1 AU and less, the solar wind turbulence 
is still developing (see, e.g., the structure function analysis of \cite{PagBal03}); thus, the 
solar wind structures that contribute to intermittency have not reached statistical stationarity. 
At larger radial distances and higher heliolatitudes, the turbulence is
decaying as there are fewer ``sources" of turbulence than
at low latitudes, like, e.g., velocity
stream shears. This hypothesis could be tested by investigation of \textit{Ulysses} data not included in this study.
Another  effect that could possibly contribute to the decrease
of intermittency with the radial distance is the inherent
expansion of the structures involved in the intermittent transfer of energy; their statistics would therefore decrease within
the ranges of scales considered in this study. Possible tests
of this hypothesis could be provided by numerical simulations
with an expanding solar wind.\\
Based on the investigations presented in this Letter, we conclude  that the evolution of 
``pure'' slow and fast solar wind turbulence beyond the ecliptic plane is probably insufficient to maintain a high level of intermittency. 
While previous analyses of the radial evolution of solar wind intermittency were based on
probability distribution functions and their moments, in this study we use the degree of
multifractalality as a key descriptor. Although the two approaches have similarities,
they do not capture exactly the same aspects of solar wind fluctuations and scale 
behavior. In a future study, we will explore the consistency between the two approaches for the same data set.

\acknowledgments
We would like to thank the SWOOPS, SWICS, and MAG instruments teams of the \textit{ULYSSES} mission for providing data. 
This work was supported by the European Community's Seventh Framework Programme ([FP7/2007-2013]) under grant agreement No. 313038/STORM.
A.W. and W.M.M. were additionally supported by the National Science Center, Poland (NCN), through grant 2014/15/B/ST9/04782. M.M.E. acknowledges support from the Romanian Ministry of National Education, CNCS, UEFISCDI, project number PN-II-ID PCE-2012-4-0418, and from the Interuniversity Attraction Poles Programme initiated by the Belgian Science Policy Office (IAP P7/08 CHARM).


\end{document}